# Studying Internal Compositions of Binary Alloy Pd-Rh Nanoparticles Using Bragg Coherent Diffraction Imaging

Tomoya Kawaguchi,[1,2] Wonsuk Cha,[3] Vitalii Latyshev,[4] Serhii Vorobiov,[4] Vladimir Komanicky,[4] and Hoydoo You[1,*]

[1]*Materials Science Division, Argonne National Laboratory, Argonne, Illinois, USA*

[2]*Institute for Materials Research, Tohoku University, Sendai, Japan*

[3]*Advanced Photon Source, Argonne National Laboratory, Argonne, Illinois, USA*

[4]*Department of Condensed Matters Physics, Safarik University, Kosice, Slovak Republic*

## ABSTRACT

Bragg coherent diffraction imaging (BCDI), the well-established technique for imaging internal strain of nanoparticles, was used to image the internal compositional distribution of binary alloys in thermal equilibrium. The images experimentally obtained for Pd-Rh alloy nanoparticles are presented and discussed. The direct correspondence between the lattice strain and the compositional deviation is discussed with the derivation of the BCDI displacement field aided by illustrations. The correspondence suggests that the longitudinal derivative of the displacement field, the strain induced by compositional heterogeneity, can be quantitatively converted to the 3D images of compositional deviation from the particle average using Vegard's law. It also suggests that the transverse derivative can be qualitatively associated with the disorder of Bragg planes. The studied Pd-Rh alloy nanoparticle exhibited internal composition heterogeneity; Rh composition tends to be high at edges and corners between facets and gradually decreased from the surface to core of the particle.

[*] Email: hyou@anl.gov
 Fax: +1-630-252-7777

# I. INTRODUCTION

The internal composition distribution of binary alloy particles is of great interest in understanding activities of metallic catalysts. It is known that the surface composition of a binary alloy particle can change with environments, e.g., reducing and oxidizing gas environments,[1] and eventually alters the catalytic activity. Therefore, it is important to have experimental techniques that can monitor *in situ* the distributions of the internal composition while the environment dynamically changes. Such a technique can advance the frontiers of alloy catalysts for various chemical and electrochemical reactions.

Bragg coherent diffraction imaging (BCDI) is a well-established X-ray technique becoming increasingly popular for imaging nanoparticles under *in-situ* conditions[2-4]. In BCDI, a 3D real space image is reconstructed from a 3D intensity distribution of a Bragg peak from a nanocrystal illuminated with a coherent X-ray beam typically available from a synchrotron source. The reconstructed real-space 3D image is composed of voxels of complex numbers. The amplitude of a voxel is proportional to the *ordered crystalline density* [5,6], while the phase is proportional to the *displacement* of the diffracting planes from the average Bragg lattice spacing. Additionally, it has been shown that the gradient of the displacement field can be associated with details of lattice strains unavailable in other experimental techniques[7-9].

A question arises whether the displacement field in nanocrystals can be applied to deduce internal composition distributions in the case of binary alloy nanocrystals in a similar manner. When two elements of an alloy nanoparticle have sufficiently different atomic numbers, e.g., Rh and Pd, the 3D image of the electron density obtained in BCDI in principle could encode the composition ratio. However, the degree of crystalline order within the particle is generally unknown, which makes it difficult to determine compositional variations from the amplitude. Furthermore, distinguishing the neighboring elements in the periodic table, e.g., Rh and Pd, is challenging. The phase of a BCDI image, on the other hand, is a reliable measure for lattice spacing with high precision as it will become clear below. In this paper, following experimental section, a 3D images obtained from a Pd-Rh alloy nanoparticle will be presented. Then the discussion of the result will include the relationship of the 3D phase image in BCDI to lattice spacing distribution and to composition distribution. The discussion will then be closed with the section of concluding remarks. Derivation of the BCDI phase and composition strains are also given in Appendices for interested readers.

## II. EXPERIMENTS

The samples of Pd-Rh alloy particles were prepared by a dewetting process. Thin films of Pd and Rh mixture were co-deposited on $SrTiO_3$ (STO) by magnetron sputtering from two targets of Pd and Rh at room temperature with discharge powers of 70 W and 75 W, respectively, to obtain 20 nm total thickness with 1:1 composition. The thickness was monitored with quartz crystal microbalance (QCM) during growth and composition was verified by an energy-dispersive x-ray (EDX) analysis. Then, the deposited Pd-Rh thin film was annealed at 1000 °C for 5 min in $N_2$ atmosphere and cooled slowly to obtain nanoparticles.

The BCDI measurements were performed at the beamline 34ID-C, Advanced photon source (APS), Argonne National Laboratory. Helium gas with 5% $O_2$ under atmospheric pressure was flowed at 310 °C to avoid palladium hydride formation. X-ray energy was 9.0 kV monochromatized by a pair of Si(111) crystals and the beam was focused to $0.6(V) \times 1(H)$ μm$^2$ at the sample position using a set of Kirkpatrick–Baez mirrors. Coherent diffraction images of cubic (111) Bragg peak was acquired by an X-ray-sensitive area detector (Medipix2/Timepix, composed of 256×256 pixels with the pixel size of 55 μm×55 μm). 3D coherent diffraction data were obtained with a rocking scan (~0.6°) by correcting the sensitivity distribution on the detector, subtracting background, eliminating alien scatterings from the other particle, and stacking the processed 2D images. 3D complex electron density distributions were reconstructed from the data by iterative phase retrieval algorithm[10] in combination with a guided algorithm.[11]

### III. RESULTS

The sample prepared by the dewetting process was examined by scanning electron microscopy (SEM). An SEM image for a small portion of the sample surface is shown in Fig. 1(a) and the diffraction pattern from one of the nanoparticles, chosen for a good signal-to-background, is shown in (b). An image of the particle measured with x-rays is not shown here because it is difficult to identify the diffracting particle without fiducial markers [12] and the image (a) represents a random ensemble of the particles.

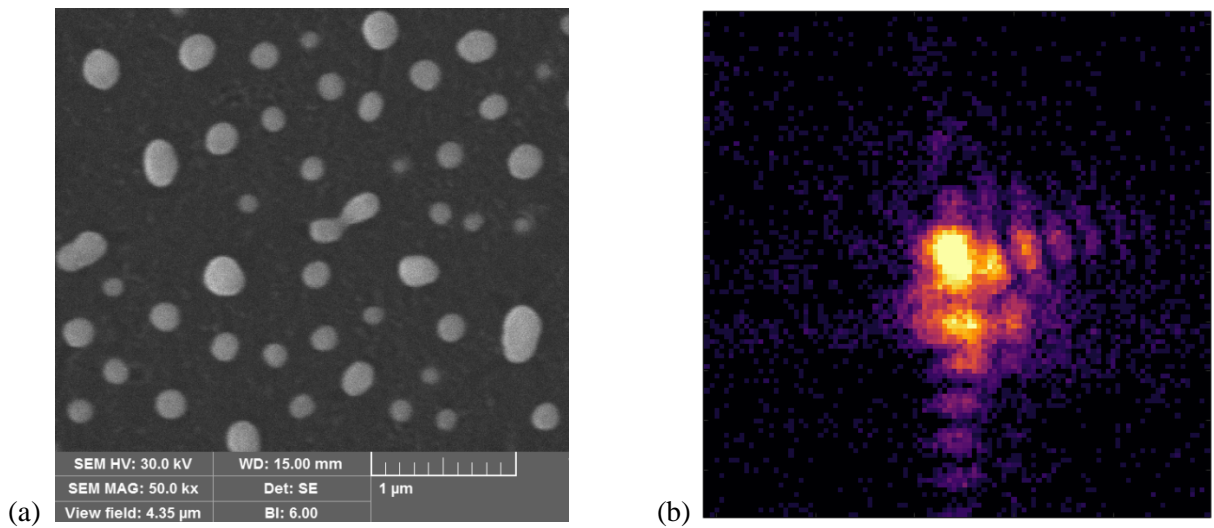

Fig. 1. (a) An SEM image of the Pd-Rh nanoparticles grown on STO substrate. Only one nanoparticle with ~200 nm size was selected for a well-defined fringe pattern and used in obtaining the diffraction patterns. (b) A typical Bragg diffraction pattern from the particle in a square root intensity scale. Sixty diffraction patterns like this were obtained in a rocking scan to form a 3D data set.

We focused on one particle in this study while we checked several particles in the course of the experiment. First, the average composition of this particle was determined from the measured $2\theta$ angle and reported lattice constants[13,14] at this temperature using Vegard's law. The composition determined in this way was $Pd_{0.22}Rh_{0.78}$. This composition was different from the 1:1 initial composition obtained by the sputtering process (Sec. II). There are two possible reasons for this discrepancy. The post annealing process for dewetting can change the composition because the vapor pressures of the elements are often different. The element with a higher vapor pressure loses the composition during the annealing. Another reason stems from the phase separation expected from the equilibrium phase diagram [15]. According to the phase diagram, $Pd_{0.5}Rh_{0.5}$ is not allowed in thermodynamic equilibrium condition. Either way, it is not surprising to find the composition quite different from the deposition ratio and investigating the reason for the discrepancy is outside the scope of this paper.

The 3D images of the amplitude and the phase for the $Pd_{0.22}Rh_{0.78}$ nanoparticle obtained in BCDI analysis are shown in Fig. 2. The shape of the particle that we studied was ~200 nm triangular prism with the thickness of ~100 nm as shown in (a). The electron density inside the particle, obtained from the amplitude part of the 3D image, is homogeneous and neither holes nor voids were observed in cross sectional images of electron density (b). The range of the displacement, obtained from the phase of the image was within $d_{111}$ and no discontinuity, indicating no detectable dislocation. Thus, we took the gradient of the displacement to obtain the longitudinal and transverse components of the compositional strain (c,e)[12] as discussed in Appendices. The longitudinal component was further converted, as outlined in Sec. IV, to the Rh composition difference (d) from the particle average.

The Rh composition tended to be higher at the corner regions formed by the facets near the surface and gradually decreased towards the core of the particle. The core of the particle have a rather uniform composition but with Rh slightly deficient compared to the overall particle average. On the other hand, the transverse term was highly concentrated at the surface and corners of the particle, while it was negligibly small inside the particle. The large transverse term at the surface can be attributed to the significant disorders in the lattice planes at the very surface possibly due to formation of the surface oxide. The formation of Rh oxides at the surface also explains the enhanced Rh composition as seen in Rh-Pt alloy systems.[12]

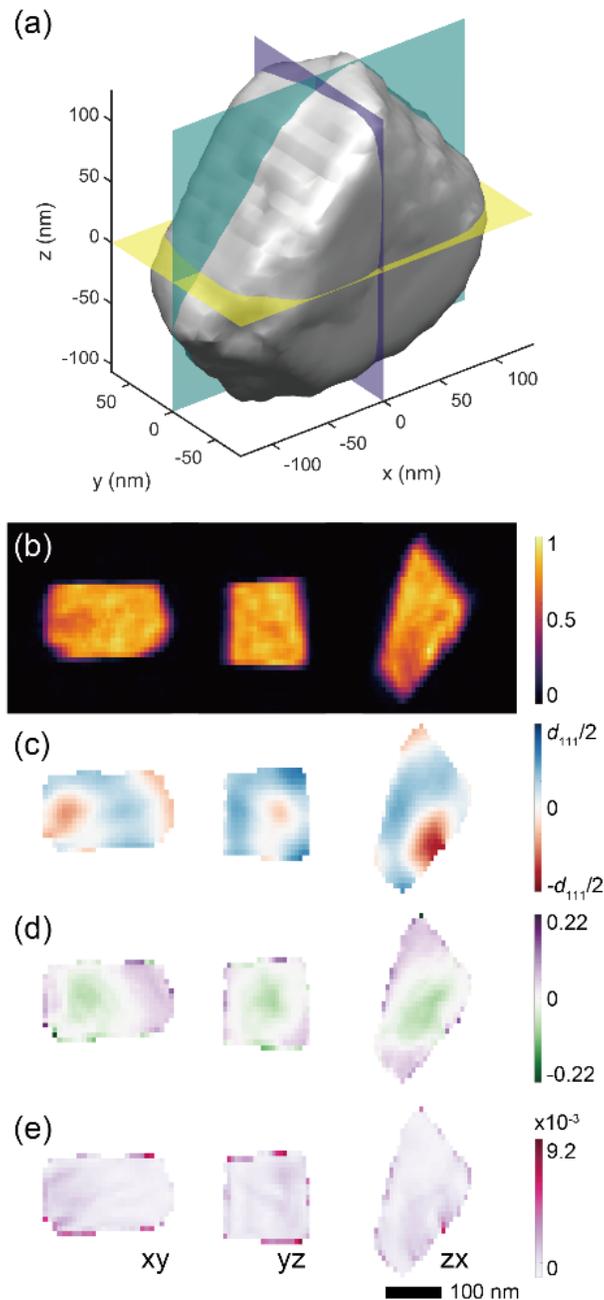

Fig. 2. (a) A 3D reconstruction of the Pd-Rh particle using BCDI. x, y, and z of the laboratory coordinate system correspond to the outboard direction of the storage ring, the vertical direction, and the incident x-ray direction, respectively. Transparent *xy, yz* and *zx* planes are used to display the cross-sectional views of (b-e). (b) electron density, (c) displacement of lattice planes (d) Rh composition ($\Delta x_{Rh}(r)$) difference form the particle average, and (e) distribution of the absolute of transverse term at *xy, yz* and *zx* planes. The voxel size was 6 nm, which is determined from the detector pixels and the angular step of the measurements.

# IV. DISCUSSION: BCDI STRUCTURE FACTOR TO COMPOSITION IMAGES

Both Pd and Rh elements have the face-centered cubic (fcc) symmetry with 2.3% difference in lattice constants (3.8907 vs. 3.8034 Å) and a Rh atom occupies ~7% smaller volume than Pd does. The phase diagram indicates that they undergo a phase separation at ~800 °C and there is no known compositional ordering [15]. While it is of scientific interest to study phase separation phenomena in nanoparticles, our focus here is to develop a BCDI methodology for a 3D internal compositional distribution of a binary alloy nanoparticle.

In the case of binary alloys like Pd-Rh, the lattice displacement field (BCDI phase – See Appendix I) can originate from two sources: compositional changes and elastic lattice strain. When the atoms are free to move at high temperatures, however, the elastic strain can be relieved or minimized by the composition redistribution as long as the two component atoms have a significant size difference and are allowed to diffuse. This is a reasonable assumption in binary alloy nanoparticles at high temperatures. For example, let us assume that a region has an elastic compressive strain because the lattice spacings are smaller than the average. In this case, the smaller atoms, Rh in this case, can easily move into the region by exchanging the positions with the larger Pd atoms. In this way, the elastic strain and total free energy are minimized. That is, the lattice strain must result all, at least mostly, from the composition differences in the absence of other driving forces. However, the other driving forces can complicate the situation. For example, an external driving force such as surface reactions of a nanoparticle alloy can significantly change the core-shell composition [12]. An internal driving force, such as spinodal decomposition that is potentially important to this Pd-Rh case, can also complicates the above assumption. However, the equilibrium spinodal decomposition minimize the total free energy by paying penalties at the sharp interfaces between the separated phases. Therefore, the internal compositions in each domain can still be imaged with the assumption of no mechanical strain, even though we will not be able to image the boundaries.

Nevertheless, as the BCDI resolution improves in future, the technique for separating the mechanical strain and compositional strain will be desirable to image the phase boundaries, for example, by carefully comparing the phase image with the amplitude image.

Under the assumption, we define the compositional strain, like the elastic strain [16], as the *longitudinal* component of the gradient of the displacement field, $u_{111}(\mathbf{r})$ as

$$s_{||} = \hat{\mathbf{Q}}_{111} s_{||}(r) = \hat{\mathbf{Q}}_{111} \cdot \nabla u_{111}(r) \qquad \text{Eq. (1)}$$

where the subscription 111 is used to emphasize the fact that the displacement of the (111) Bragg planes. A simple example for the relationship of the 1D displacement field to compositional strain and to the lattice spacing deviation from the average is illustrated in Fig. 3(a). It shows a region with lattice planes at the position **r**, i.e., lattice planes are constant transverse directions as much as the size of voxel[2] in the image. In this example, the lattice spacing is uniform but smaller in the left side or larger in the right side than the average (the dashed lines). As we can, the displacement, the difference between the dashed lines and solid lines, increases in both directions. However, the derivatives of the displacement have two difference signs (+ in the right and – in the left side), indicating the contracted and expanded regions, respectively.

The conversion to Rh composition from the longitudinal strain, $s_{//}(\mathbf{r}) \equiv s_{111}(\mathbf{r})$, is performed by using a relationship between the composition and lattice spacing. According to the Vegard's law, the Rh composition at each voxel is calculated as $x_{\text{Rh}}(r) = \bar{x}_{\text{Rh}} + \Delta x_{\text{Rh}}(r)$. $\bar{x}_{\text{Rh}}$, the average composition determined from $2\theta$ of the Bragg peak position, is $(d_{\text{Pd}} - d_{\text{ave}})/(d_{\text{Pd}} - d_{\text{Rh}})$ where $d_{\text{ave}}$ is the lattice spacing calculated from the Bragg peak position and $\Delta x_{\text{Rh}}(\mathbf{r})$, the local composition difference from the average, is $-\frac{s_{111}(r) d_{\text{ave}}}{d_{\text{Pd}} - d_{\text{Rh}}}$ where $d_{\text{Pd}}$ and $d_{\text{Rh}}$ are 111 lattice spacing of pure Pd and pure Rh at the measurement temperature, respectively.[13] $\Delta x_{\text{Rh}}(r)$ is displayed in Fig. 2(d).

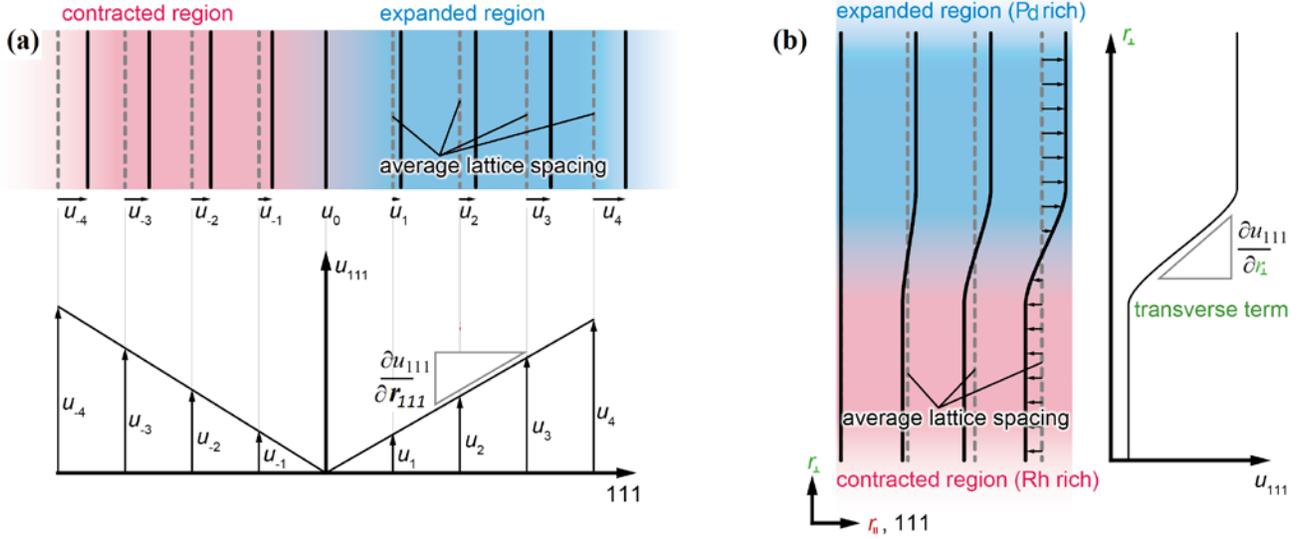

Fig. 3. Schematics of the lattice displacements at a boundary where lattice spacings change in the (a) longitudinal and (b) transverse directions. Note the seamless boundary in (a) and the kink boundary in (b) between the regions of the expanded and contracted spacings. In (a), the lattice spacings of the two regions are obtained from the longitudinal derivatives of the displacement. In (b), the transverse derivatives are non-zero only in the boundaries where the lattice spacings change.

Now, we define the "transverse term" that is a vector as,

$$\boldsymbol{s}_\perp(\boldsymbol{r}) = \nabla u_{111}(\boldsymbol{r}) - s_\parallel \cdot \widehat{\boldsymbol{Q}}_{111} \qquad \text{Eq. (2)}$$

which is a simply the remainder vector of the gradient after the longitudinal component is subtracted. A simple case, when the y-axis is set to the direction of the transverse vector direction, is illustrated in Fig. 3(b). The meaning of the transverse term is not as clear as the longitudinal term because of the vectorial nature. However, with the help of the illustration for the simple case, we identify the transverse term to be qualitatively proportional to the degree of disorder in the lattice spacings. That is, the transverse term is non-zero when the lattice spacing changes in the transverse direction as illustrated or the lattice planes are *disordered*. The transverse term for the Pd-Rh is displayed in Fig. 2(e). The image indicate that only the extreme edges and corners appear disordered and the rest of the crystal body appears well ordered.

In summary, the longitudinal derivative is directly proportional to the lattice spacing changes and the transverse derivative is qualitatively proportional to the lattice spacing disorder. Although BCDI does not have the atomic level resolution, the 3D voxel-level phase can be used as an approximation to the atomic level lattice spacing using their derivatives along the 111 direction, providing the information about the lattice expansion/contraction and eventually the local composition. In this manner, BCDI offers the possibility of measuring the internal dynamics of the compositional redistribution of a nanoparticle *in situ* in reactive environments, which is important for a deeper understanding of the catalytic behavior of bimetallic nanoparticles and for further developing alloy catalysts for enhanced activity and selectivity[17].

# V. CONCLUSION

The correspondence between the lattice strain obtained from the BCDI and composition deviation was discussed. The longitudinal derivative of displacement field, i.e., phase image obtained in BCDI, indicates the compositional heterogeneity while the transverse term implies the disorder of Bragg planes. The Pd-Rh binary alloy particle was exemplified for the internal composition analysis using the displacement of BCDI, in which Pd and Rh are the neighboring elements and too similar to distinguish only from the electron density. The edge and surface of the particle tend to be Rh-rich and the Rh composition gradually and monotonically decrease from the surface to the core of the particle, which would be related to the shape dependence of the catalytic activity.

# ACKNOWLEDGEMENT

The work at Argonne (TK HY) was supported by the U.S. Department of Energy (DOE), Office of Basic Energy Science (BES), Materials Sciences and Engineering Division, and use of the APS and work at APS (WC), by DOE BES Scientific User Facilities Division, under Contract No. DE-AC02-06CH11357. The work at Safarik has been supported by grant VEGA No. 1/0204/18, the grant of the Slovak Research and Development Agency under the contract No. APVV-17-0059 and by the ERDF EU grant under the contract No. ITMS26220120047. One of the authors (TK) thanks the Japanese Society for the Promotion of Science (JSPS) for JSPS Postdoctoral Fellowships for Research Abroad.**APPENDIX A: STRUCTURE FACTOR OF BRAGG REFLECTION IN BCDI**

We provide a simple writing of the BCDI structure factor in a few steps of equations by focusing only on the structure factor for coherent x-ray diffraction using a Bragg peak based on the published results [9]. Let us define the position of an atom by $\mathbf{r}_j$ in real space as usual. Now, we rewrite $\mathbf{r}_j = \mathbf{r}_j^G + \mathbf{u}_j$ as the sum of an ideal lattice position ($\mathbf{r}_j^G$), e.g., one of fcc lattice positions, and displacement ($\mathbf{u}_j$) away from the ideal position. Similarly, we define $\mathbf{Q} = \mathbf{q} + \mathbf{G}$ as the sum of the wave vector deviation ($\mathbf{q}$) and the wave vector for Bragg peak ($\mathbf{G}$). Then, the scattering amplitude, $A(\mathbf{Q})$, can be written in a lattice and converted to a continuum limit as,

$$\begin{aligned}
A(\mathbf{Q}) &= F(\mathbf{Q}) \sum_{j}^{\infty} s(\mathbf{r}_j) \exp(-i\mathbf{Q} \cdot \mathbf{r}_j) \\
&= F(\mathbf{Q}) \sum_{j}^{\infty} s(\mathbf{r}_j) \exp\left\{-i\left(\mathbf{q} \cdot \mathbf{r}_j^G + \mathbf{q} \cdot \mathbf{u}_j + \mathbf{G} \cdot \mathbf{r}_j^G + \mathbf{G} \cdot \mathbf{u}_j\right)\right\} \\
&\simeq F(\mathbf{Q}) \sum_{j}^{\infty} s(\mathbf{r}_j^G) \exp\{-i\mathbf{G} \cdot \mathbf{u}_j\} \exp\{-i\mathbf{q} \cdot \mathbf{r}_j^G\} \\
&= F(\mathbf{Q}) \int d\mathbf{r} [s(\mathbf{r}) \exp\{-i\mathbf{Q} \cdot \mathbf{u}(\mathbf{r})\}] \exp\{-i\mathbf{q} \cdot \mathbf{r}\}
\end{aligned} \qquad \text{Eq. (3)}$$

where $s(\mathbf{r}_j)$ is the shape function. It is zero outside the nanocrystal and depends on electron density at $\mathbf{r}_j$. $F(\mathbf{Q})$ is the atomic form factor averaged for two alloy elements providing only $|Q|$ dependence, which is essentially constant in BCDI and its $\mathbf{r}$ dependence is included in $s(\mathbf{r}_j)$. Here, $\mathbf{q}.\mathbf{u}_j$ is negligibly small, at least we assume, because $|\mathbf{q}|$ and $|\mathbf{u}|$ both are very small compared to $|\mathbf{G}|$ and $|\mathbf{r}^G|$. Then, $\mathbf{G}.\mathbf{r}_j^G$ is a multiple of $2\pi$ by definition and can be ignored. Therefore, the final form of the scattering amplitude is a simple Fourier transformation of the complex object, $s(\mathbf{r}_j)\exp\{-i\mathbf{Q}.\mathbf{u}(\mathbf{r})\}$, with the shape amplitude distribution, $s(\mathbf{r}_j^G)$, and the phase distribution, $\mathbf{Q}.\mathbf{u}(\mathbf{r})$, as used in BCDI.

## APPENDIX B: DISPLACEMENT, STRAIN, AND COMPOSITION

Let us examine the definition of **G** used in the previous section for an alloy. In the example of the Pd-Rh alloy, $|\mathbf{G}_{Pd}| < |\mathbf{G}| < |\mathbf{G}_{Rh}|$ and $|\mathbf{G}|$ depends on the composition. This will be a trivial result for a bulk alloy where $|\mathbf{G}|$ will be determined by the average composition and is the same everywhere in the bulk. However, in a nanocrystal with a large surface to bulk ratio, the presence of surface and surface reaction with gas alone can make the situation complicated and interesting.[12]

The displacement field, $u(\mathbf{r})$, can be constructed in 3D from the phase of a BCDI image. Then, $u(\mathbf{r})$ can be converted to $x_{Rh}(\mathbf{r})$ the compositional image. The following simple equations and the illustrations shown in Fig. 3 can be used to explain how a strain field is related to the compositional distribution. Using a cylindrical coordinate system with the $z$-direction along the measured **G** direction, the diffraction planes are defined with a discrete lattice, $z_j$, and a continuum in-plane vector, $\boldsymbol{\rho} = (\rho, \varphi)$. Then a position, **r** near the $j^{th}$ layer is

$$\begin{aligned} \mathbf{r}_j &= \boldsymbol{\rho} + z_j \hat{\mathbf{z}} \\ &= \boldsymbol{\rho} + [j\bar{c} + u(\boldsymbol{\rho}, z_j)]\hat{\mathbf{z}} \end{aligned}$$

Eq. (4)

where the displacement, $u(\mathbf{r}) = u(\boldsymbol{\rho}, z_j)$, is generally a fraction of the particle-average lattice spacing $\bar{c}$. Then, the lattice spacing, $c(\boldsymbol{\rho}, z_j)$, between the $j^{th}$ and $(j+1)^{th}$ layers then becomes

$$\begin{aligned} c(\boldsymbol{\rho}, z_j) &= \mathbf{r}_{j+1} - \mathbf{r}_j \\ &= \bar{c} + u(\boldsymbol{\rho}, z_{j+1}) - u(\boldsymbol{\rho}, z_j) \\ &\simeq \bar{c} + \bar{c}\frac{\partial u(\boldsymbol{\rho}, z)}{\partial z}\bigg|_{z=z_j} \end{aligned}$$

Eq. (5)

Here, we can see that the lattice spacing at a given vector position, **r**, is given by the one-dimensional derivative of the displacement field, u(**r**), along the direction of **G**. A simple case of this term is illustrated in Fig. 3(a). It shows a region with lattice planes at the position **r**, i.e., lattice planes are constant transverse directions, at least over the size of voxel[2] in the image.

Let us consider 3D derivatives instead of 1D derivative. In 3D, the displacement difference, $u(\mathbf{r}+\Delta\mathbf{r}) - u(\mathbf{r})$, between two neighboring positions separated by $\Delta\mathbf{r}$ can be expanded to using the above definition of the lattice spacing,

$$u(\mathbf{r} + \Delta\mathbf{r}) - u(\mathbf{r}) \simeq \Delta\mathbf{r} \cdot \nabla u(\mathbf{r})$$
$$= \left[ \Delta x \frac{\partial u(\rho, z)}{\partial x} + \Delta y \frac{\partial u(\rho, z)}{\partial y} \right] + \frac{\Delta z}{\bar{c}}[c(\rho, z) - \bar{c}] \qquad \text{Eq. (6)}$$

Here, the second term, the *longitudinal* term along the $z$-direction, was discussed in the 1D derivative for the lattice spacing deviation from the average and illustrated in Fig. 3(a).

We define the longitudinal term as

$$s_{||} = \widehat{\mathbf{Q}}_{111} s_{||}(\mathbf{r}) = \widehat{\mathbf{Q}}_{111} \cdot \nabla u_{111}(\mathbf{r}) \qquad \text{Eq. (7)}$$

where $u_{111}(\mathbf{r})$ is used instead of $u(\mathbf{r})$ to emphasize that the displacement was measured on a (111) Bragg peak. The term in the bracket is the remainder of the phase difference and we define the "transverse term" that is a vector as,

$$\mathbf{s}_\perp = (\mathbf{r}) = \nabla u_{111}(\mathbf{r}) - s_{||} s_{||} \cdot \widehat{\mathbf{Q}}_{111} \qquad \text{Eq. (8)}$$

A simple case, where one of the terms in the bracket is zero, is illustrated in Fig. 3(b). The transverse term is qualitatively proportional to the degree of disorder in the lattice spacings. That is, the transverse term is non-zero when the lattice spacing changes in the transverse direction as illustrated or the lattice planes are *disordered*.